\documentclass[preprint,showpacs,amsmath,amssymb,aps,prd]{revtex4}

\usepackage{graphicx}

\usepackage{amsmath}

 \newcommand{\ds}{\displaystyle}

\begin{document}

\title{Gravitational acceleration in a class of geometric sigma models}

\author{Milovan Vasili\'c} \email{mvasilic@ipb.ac.rs}
\affiliation{Institute of Physics, University of Belgrade, P.O.Box 57,
11001 Belgrade, Serbia}

\date{\today}

\begin{abstract}
In this work, I examine spherically symmetric solutions in geometric
sigma models with four scalar fields. This class of models turns out
to be a subclass of the wider class of scalar-vector-tensor theories
of gravity. The purpose of the present study is to examine how the
additional four degrees of freedom modify Newtonian gravitational
acceleration. I have restricted my considerations to pointlike sources
in de Sitter background. The resulting gravitational acceleration has
the form of a power series, with four major terms standing out. The
first and the second are the familiar Newtonian and MOND terms, which
dominate at short distances. The third term is dominant at large
distances. It is the $\Lambda$CDM term responsible for the accelerated
expansion of the Universe. Finally, the fourth term provides an extra
repulsive acceleration that grows exponentially fast with distance.
This term becomes significant only at extremely large distances that
go beyond the observable Universe. As for the time dependence of the
calculated gravitational acceleration, it turns out to have
nontrivial, oscillatory character.
\end{abstract}


\maketitle

\section{Introduction}\label{Sec1}

In the existing literature, the subjects of the early and the late
time cosmologies are almost exclusively addressed separately. Indeed,
the present epoch is commonly described by the standard cosmological
model ($\Lambda$CDM), which makes no predictions concerning the early
Universe. In particular, it lacks inflation, which is believed to
correctly describe the early Universe. On the other hand, the
inflationary models that one encounters in scientific literature are
hardly ever checked for their influence on the small scale problems of
the present epoch. For example, the known problem of flat galactic
curves \cite{1,2,3,4,5,6,7,8} may well be connected to the
modification of gravity brought by the inflationary models. Thus,
before making an ad hoc modification of gravity, one is advised to
first examine the modifications found in the existing cosmological
models.

In what follows, I shall examine a class of geometric sigma models
with four scalar fields. These models have first been proposed in
Ref.\ \cite{16} in the context of fermionic excitations of flat
geometry. In Ref.\ \cite{18} they are used for the construction of
various inflationary and bouncing cosmologies. It has been shown that
the resulting cosmologies have everywhere regular and stable
backgrounds irrespective of their specific types. In particular, small
metric perturbations are demonstrated to regularly pass through the
bounce. By inspecting the particle spectrum of these models, they are
shown to belong to a wider class of scalar-vector-tensor theories of
gravity. My primary motivation is to find out if these theories can be
as successful in explaining the late time behavior of the Universe as
they are successful in explaining the early Universe. In particular, I
want to calculate the modified gravitational acceleration and see if it can
fit the observation. To this end, I shall consider a pointlike source,
and calculate spherically symmetric metric far from it.

The results of the paper are summarized as follows. The needed
spherically symmetric solution is found for the whole class of
considered geometric sigma models. The solution is obtained in a weak
field approximation, and in the form of a power series. As it turns
out, the corresponding coefficients are subject to a set of well
defined recurrent relations. These recurrent relations are solved in
de Sitter background, which is commonly assumed to characterize the
present epoch. This way, the obtained late time behavior of the
gravitational acceleration is shown to hold true for all the models
with $\Lambda$CDM limit. The integration constants are determined from
the requirement that gravitational acceleration reproduces Milgrom's
modified Newtonian dynamics (MOND) at short distances, and the
familiar $\Lambda$CDM behavior at large distances. The resulting
expression is a sum of four major contributions, which one by one,
become dominant as the distance from the gravitational source grows.
At short distances, the dominant contribution comes from the Newtonian
term. Then, at longer distances, the dominant role is taken by the
familiar MOND term. At even larger distances, the leading role is
carried by the $\Lambda$CDM term which is responsible for the
accelerated expansion of the Universe. Finally, the fourth term
provides an extra repulsive acceleration that grows exponentially fast
with distance. This term becomes significant only at extremely large
distances that go beyond the observable Universe. As such, it can be
neglected in practically all astronomical measurements. As the final
achievement of this work, let me mention time dependence of the
gravitational acceleration. It is shown that gravitational
acceleration of the pointlike source has oscillatory dependence on
cosmic time. In particular, the gravitational acceleration that is
attractive at the present time could have been repulsive at earlier
times. 

The results obtained in this paper should be confronted with results
of similar considerations in literature. Plenty of modified gravity
theories considered in literature predict corrections to the Newtonian
gravitational force that can explain the unexpected astronomical data.
For example, the authors of Refs.\ \cite{18x} and \cite{18y} consider
a class of $f(R)$ theories of gravity, and succeed in recovering the
observed behavior of many spiral and elliptical galaxies. In
particular, the $f(R)\propto R^{3/2}$ theory is shown to lead to the
familiar MOND behavior \cite{18y}. However, all these results are
obtained by considering a static metric in a flat background. As a
consequence, the obtained gravitational force is necessarily time
independent. This should be confronted with the present paper, where
the considered pointlike source is placed in a nontrivial cosmic
background, and the metric ansatz is not static. As a consequence, the
obtained gravitational force has a nontrivial time dependence.

The layout of the paper is as follows. In Sec.\ \ref{Sec2}, a precise
definition of the class of models to be considered is given. The very
construction of geometric sigma models is only briefly recapitulated.
In Sec.\ \ref{Sec3}, spherically symmetric ansatz is applied to field
equations. The solution is obtained in a weak field approximation, and
in the form of a power series. The resulting expression holds true for
any choice of the scale factor $a$, and the potential $W$. In Sec.\
\ref{Sec4}, a particularly simple choice of $a$ and $W$ has been made.
Specifically, the background metric that defines the model is chosen
to be of de Sitter type. This choice is in agreement with the common
belief that whatever type of the Universe is considered, its late time
behavior should be that of the $\Lambda$CDM model. In Sec.\
\ref{Sec5}, the nonrelativistic formula for gravitational acceleration
is derived. The result is compared with MOND and $\Lambda$CDM
predictions. Sec.\ \ref{Sec6} is devoted to concluding remarks.

My conventions are as follows. Indexes $\mu$, $\nu$, ... and $i$, $j$,
... from the middle of alphabet take values $0,1,2,3$. Indexes
$\alpha$, $\beta$, ... and $a$, $b$, ... from the beginning of
alphabet take values $1,2,3$. Spacetime coordinates are denoted by
$x^{\mu}$, ordinary differentiation uses comma ($X_{,\,\mu} \equiv
\partial_{\mu} X$), and covariant differentiation uses semicolon
($X_{;\,\mu}\equiv \nabla_{\mu}X$). Repeated indexes denote summation:
$X_{\alpha\alpha} \equiv X_{11} + X_{22} + X_{33}$. Signature of the
$4$-metric $g_{\mu\nu}$ is $(-,+,+,+)$, and curvature tensor is
defined as $R^{\mu}{}_{\nu\lambda\rho} \equiv \partial_{\lambda}
\Gamma^{\mu}{}_{\nu\rho} - \partial_{\rho}
\Gamma^{\mu}{}_{\nu\lambda}+\Gamma^{\mu}{}_{\sigma\lambda}
\Gamma^{\sigma}{}_{\nu\rho} - \Gamma^{\mu}{}_{\sigma\rho}
\Gamma^{\sigma}{}_{\nu\lambda}$. Throughout the paper, the natural
units $c=\hbar=1$ are used.

\section{Geometric sigma models}\label{Sec2}

The model considered in this paper belongs to the class of geometric
sigma models, originally defined in Ref.\ \cite{16}. The main feature
of every geometric sigma model is that it is defined by associating
action functional with a fixed, freely chosen metric
$g_{\mu\nu}^{(o)}(x)$. The action has the form
\begin{equation} \label{1} 
I_g = \frac{1}{2\kappa}\int d^4x\sqrt{-g}\left[ R -
F_{ij}(\phi)\phi^i_{,\mu}\phi^{j,\mu} - V(\phi) \right] ,
\end{equation}
where $F_{ij}(\phi)$ and $V(\phi)$ are target metric and potential of
four scalar fields $\phi^i(x)$. The constant $\kappa \equiv 8\pi G$
stands for the gravitational coupling constant. The target metric
$F_{ij}(\phi)$ is constructed by replacing $x^i$ with $\phi^i$ in the
expression
\begin{equation} \label{2} 
F_{ij}(x)\equiv R_{ij}^{(o)}(x) - \frac{1}{2} V(x) g_{ij}^{(o)}(x) \,, 
\end{equation}
where $R_{\mu\nu}^{(o)}(x)$ is Ricci tensor for the metric
$g_{\mu\nu}^{(o)}(x)$. The same replacement in an arbitrary function
$V(x)$ defines the potential $V(\phi)$. This construction guarantees that
\begin{equation} \label{3} 
\phi^i = x^i \,, \quad g_{\mu\nu} = g_{\mu\nu}^{(o)} 
\end{equation} 
is a solution of the field equations defined by Eq.\ (\ref{1}). In
what follows, the solution Eq.\ (\ref{3}) will be referred to as {\it
vacuum}. It is seen that physics of small perturbations of this vacuum
allows the gauge condition 
\begin{equation} \label{3.2} 
\phi^i(x) = x^i \,.
\end{equation} 
Gauge fixed field equations employ the metric alone, and read
\begin{equation} \label{3.5}
R_{\mu\nu} = R_{\mu\nu}^{(o)}(x) + \frac{1}{2} V(x)
\left( g_{\mu\nu} - g_{\mu\nu}^{(o)} \right) .
\end{equation}
In what follows, I shall be interested in how a pointlike source
deforms the surrounding empty background. Thus, the needed field
equations will be the matter free equations (\ref{3.5}).

The cosmological background I choose to work with is the background
metric
\begin{equation} \label{4} 
ds^2 = -dt^2 + a^2(t)\left( dx^2 + dy^2 + dz^2 \right) .
\end{equation} 
It defines the vacuum metric $g_{\mu\nu}^{(o)}$. The model itself is
defined by determining $V(x)$ and $F_{ij}(x)$. For the vacuum metric
Eq.\ (\ref{4}), one finds
\begin{equation}\label{5} 
F_{00} = W - 2\dot H \,, \quad F_{0b} = 0 \,,\quad
F_{ab}=-a^2\,W \delta_{ab} \,, 
\end{equation} 
where $H\equiv \dot a/a$ is the Hubble parameter, and $W$ is defined
by
\begin{equation} \label{6} 
V \equiv 2\left( W + \dot H + 3H^2 \right) . 
\end{equation} 
The ``dot'' denotes time derivative. The target metric $F_{ij}(\phi)$
and the potential $V(\phi)$ are obtained by the substitution $x^i \to
\phi^i$ in $F_{ij}(x)$ and $V(x)$. For the time being, the scale
factor $a(t)$, and the potential $W(t)$ are kept unspecified.

In what follows, the most general case $W\neq 0$ will be considered.
This is motivated by the failure of Ref.\ \cite{18a}, where $W=0$ case
was studied, to provide an acceptable explanation of flat galactic
curves. The $W\neq 0$ geometric sigma models have extensively been
studied in Ref.\ \cite{18}. There, they were used for the construction
of various inflationary and bouncing cosmologies. It has been shown
that the resulting cosmologies have everywhere regular and stable
backgrounds irrespective of their specific types. The necessary
conditions for proving regularity and stability have been shown to
read
\begin{equation} \label{6.2} 
W<0 \,, \quad  F_{00} < 0 \,. 
\end{equation} 
If, in addition, one makes the choice
\begin{equation} \label{6.4} 
W = - \frac{\omega^2}{a^2} \,,
\end{equation} 
where $\omega$ is a constant with the dimension of mass, the
regularity and stability are guaranteed. This choice of $W$ is not
unique, but is certainly sufficient to make the theory everywhere well
defined. By inspecting the particle spectrum, these models have been
shown to belong to a wider class of scalar-vector-tensor theories of
gravity. All their modes are massive. In particular, the graviton mass
is
\begin{equation} \label{6.6} 
m_g = \sqrt{-2W} \,.
\end{equation} 
The good thing about this is that, in most physically relevant
situations, the value of $m_g$ stays below its experimental bound.
This makes the described class of geometric sigma models 
physically liable. My primary motivation in this paper is to
find out if these theories can be as successful in explaining the
small scale problems of the present epoch as they are successful in
explaining the early Universe. In particular, I want to calculate the
gravitational acceleration of a pointlike source in a nontrivial cosmic background.

\section{Field equations}\label{Sec3}

In what follows, matter fields are assumed to be localized in a point,
which I choose to be $\vec x = 0$. Then, the field equations in the
region $\vec x \neq 0$ reduce to those obtained from the geometric
action Eq.\ (\ref{1}). In the gauge Eq.\ (\ref{3.2}), the field
equations reduce to Eq.\ (\ref{3.5}), and possess the vacuum solution
$g_{\mu\nu} = g_{\mu\nu}^{(o)}$. What I am interested in are
spherically symmetric deviations from this vacuum, caused by the
presence of a massive particle in $\vec x = 0$. It is important to
emphasize that the gauge Eq.\ (\ref{3.2}) leaves us with no residual
gauge symmetry. Thus, no further gauge fixings are possible.

The most general spherically symmetric metric in the gauge Eq.\
(\ref{3.2}) has the form
\begin{equation} \label{7}
g_{00} = \mu \,, \quad
g_{0\alpha} = \nu \frac{x_{\alpha}}{r} \,, \quad 
g_{\alpha\beta} = \lambda P^{\parallel}_{\alpha\beta} +
\rho P^{\perp}_{\alpha\beta} \,,
\end{equation}
where $P^{\parallel}_{\alpha\beta}$ and $P^{\perp}_{\alpha\beta}$ are
parallel and orthogonal projectors on $\vec x$,
\begin{equation} \label{8}
P^{\parallel}_{\alpha\beta} \equiv 
\frac{x^{\alpha}x^{\beta}}{r^2}  \,,   \quad
P^{\perp}_{\alpha\beta} \equiv \delta_{\alpha\beta} -
\frac{x^{\alpha}x^{\beta}}{r^2} \,,
\end{equation}
and $\mu$, $\nu$, $\lambda$, $\rho$ are functions of $r$ and $t$,
only. The radius $r$ is defined by $r^2 \equiv x^2+y^2+z^2$. The field
equations (\ref{3.5}) are straightforwardly expressed in terms of
$\mu$, $\nu$, $\lambda$, $\rho$, the scale factor $a$, and the
potential $W$. In what follows, I shall use a weak field
approximation, because the nonperturbative equations turn out to be too
complicated for me to solve. Thus, I define the decomposition
\begin{equation}\label{9}
\begin{array}{rl}
&\ds \mu \equiv -1+\mu_1 \,, \quad  \nu \equiv \nu_1 \,,    \\ [0.5ex]
&\ds \lambda \equiv a^2 \left(1+\lambda_1\right) , \quad          
\rho \equiv a^2 \left(1+\rho_1\right) .                                \\ [0.5ex]
\end{array}
\end{equation}
The new fields $\mu_1$, $\nu_1$, $\lambda_1$, $\rho_1$ are assumed to
be small, so that quadratic and higher order terms can be neglected.
After a lengthy calculation, the linearized field equations are
brought to the form
\begin{subequations}\label{10}
\begin{equation}\label{10a}
\begin{array}{rl}
&\ds \frac{1}{r}\left[\lambda_1 - 
\left(r\rho_1\right)'\right]_{,0} -
H\mu'_1 - W \nu_1 = {\cal O}_2 \,,                                      \\ [0.8ex]
\end{array}
\end{equation} 
\begin{equation}\label{10b}
\begin{array}{rl}
&\ds a^2\left[2H\big(\dot\lambda_1+2\dot\rho_1\big) -
W \big(\lambda_1+2\rho_1+\mu_1\big)\right] +                \\
&\ds \frac{2}{r}\left[\lambda_1-\left(r\rho_1\right)'\right]' +
\frac{2}{r^2}\left[\lambda_1-\left(r\rho_1\right)'\right] +     \\  
&\ds 2a^2\big(\dot H+3H^2\big)\mu_1 -
4H\Big(\nu'_1+\frac{2}{r}\nu_1\Big) = {\cal O}_2 \,,            \\ [0.8ex]
\end{array}
\end{equation}
\begin{equation}\label{10c}
\begin{array}{rl}
&\ds a^2\left[\big(\ddot\lambda_1-\ddot\rho_1\big) +
3H\big(\dot\lambda_1-\dot\rho_1\big) -
2W\big(\lambda_1-\rho_1\big)\right] +                          \\
&\ds \frac{1}{r}\left[\lambda_1 -
\left(r\rho_1\right)'\right]'-\frac{2}{r^2} \left[\lambda_1 -
\left(r\rho_1\right)'\right]+\mu''_1 - \frac{1}{r}\mu'_1 -   \\
&\ds 2\Big(\dot\nu'_1-\frac{1}{r}\dot\nu_1\Big) -
2H\Big(\nu'_1-\frac{1}{r}\nu_1\Big) = {\cal O}_2 \,,      \\ [0.8ex]
\end{array}
\end{equation}
\begin{equation}\label{10d}
\begin{array}{rl}
&\ds a^2F_{00}\big(\dot\lambda_1 + 
2\dot\rho_1 + \dot\mu_1\big) +                                  
a^2 W K \big(\lambda_1 + 2\rho_1\big) +                        \\
&\ds a^2\Big[\dot F_{00} + 6HF_{00}\Big]\mu_1 -                         
2F_{00}\Big(\nu'_1+\frac{2}{r}\nu_1\Big) = {\cal O}_2 \,,

\end{array}
\end{equation}
\end{subequations}
where ${\cal O}_2$ stands for quadratic and higher order terms. To
remind you, the notation $F_{00} \equiv W-2\dot H$ has already been
used in Eq.\ (\ref{5}) where it denoted the kinetic term of the model
Lagrangian. The new notation
\begin{equation} \label{10.5}
K \equiv 2H + \frac{\dot W}{W} \,,
\end{equation}
on the other hand, is introduced for mere convenience.

The solution of Eqs.\ (\ref{10}) is searched for in the form of a power
series. Specifically, I use the decomposition
\begin{equation}\label{11}
\begin{array}{rl}
&\ds \mu_1 = \sum_{n=-\infty}^{\infty}\alpha_n r^{n} , \quad    
\nu_1 = \sum_{n=-\infty}^{\infty}\delta_n r^{n} ,                         \\ [1.5ex]
&\ds \lambda_1 = \sum_{n=-\infty}^{\infty}\beta_n r^{n} , \quad     
\rho_1 = \sum_{n=-\infty}^{\infty}\gamma_n r^{n} ,
\end{array}
\end{equation}
where $\alpha_n(t)$, $\beta_n(t)$, $\gamma_n(t)$, $\delta_n(t)$ are
time dependent coefficients. The substitution of Eq.\ (\ref{11}) into
Eqs.\ (\ref{10}) yields a set of ordinary differential equations. 
Using the shorthand notation
\begin{subequations}\label{12}
\begin{equation}\label{12a}
A_n \equiv 
\dot\beta_{n+1} - (n+2)\dot\gamma_{n+1} - 
(n+1) H\alpha_{n+1} - W\delta_n \,, 
\end{equation}
\begin{equation}\label{12b}
\begin{array}{rl}
B_n \equiv 
&\ds 2a^2 H \big( \dot\beta_n + 2\dot\gamma_n\big) -
a^2 W\big( \beta_n + 2\gamma_n+\alpha_n\big) +               \\
&\ds 2(n+3)\big[\beta_{n+2} -
(n+3)\gamma_{n+2}-2H\delta_{n+1}\big] +                        \\ 
&\ds 2a^2 \big(\dot H+3H^2\big)\alpha_n  \,,       
\end{array}
\end{equation}
\begin{equation}\label{12c}
\begin{array}{rl}
C_n \equiv 
&\ds a^2 \Big[ \ddot\beta_n - \ddot\gamma_n + 
3H \big( \dot\beta_n - \dot\gamma_n \big) -
2W\big(\beta_n - \gamma_n\big) \Big] +                               \\
&\ds n\big[ \beta_{n+2} - (n+3)\gamma_{n+2} + 
(n+2)\alpha_{n+2} \big] -                                                    \\
&\ds 2n\big(\dot\delta_{n+1}+H\delta_{n+1}\big) \,,
\end{array}
\end{equation}
\begin{equation}\label{12d}
\begin{array}{rl}
D_n \equiv 
&\ds a^2F_{00}\big(\dot\beta_n + 
2\dot\gamma_n + \dot\alpha_n\big) +                                 
a^2 W K \big( \beta_n + 2\gamma_n \big) +                       \\
&\ds a^2 \Big[ \dot F_{00} + 
6H F_{00} \Big] \alpha_n -                          
2(n+3)F_{00}\,\delta_{n+1} \,,
\end{array}
\end{equation}
\end{subequations}
this set of ordinary differential equations is written as
\begin{equation}\label{13}
A_n = B_n = C_n = D_n =0 
\end{equation}
for all $n\in\mathbb{Z}$. It is seen that only $C_n = 0$ contain
second order time derivatives. In what follows, I shall get rid of
these by considering the identity
\begin{equation}\label{14}
\begin{array}{rl}
&\ds 6H\big(a^3 A_{n-1}\big)_{\!,0} -
2n(n+3)a A_{n+1} + n\big(a B_n\big)_{\!,0} -                    \\
&\ds 2(n+3)aH C_n - n a D_n  -          
3a^3 HW E_n \equiv 0 \,,
\end{array}
\end{equation} 
where $E_n$ is short for
\begin{equation}\label{15}
\begin{array}{rl}
E_n \equiv 
&\ds n\alpha_n + 
\big(n+4\big)\beta_n -
2\big(n+2\big)\gamma_n -                                                    \\
&\ds 2\left[\dot\delta_{n-1} +
\big(H+K\big)\delta_{n-1}\right] .
\end{array}
\end{equation} 
Obviously, the new equations $E_n = 0$ can replace $C_n = 0$ whenever
$n\neq -3$. This way, the only remaining second order differential
equation is $C_{-3} = 0$. A proper rearrangement of the equations $A_n
= B_n = D_n = E_n =0$ finally yields the needed recurrent relations:
\begin{widetext}
\begin{subequations}\label{16}
\begin{equation}\label{16a}
\beta_{n+2} - (n+3) \gamma_{n+2} = 
\frac{a^2 H}{n+3} \bigg[ \dot\alpha_n + 
\bigg( 3H + \frac{F_{00}}{2H} + 
\frac{\dot F_{00}}{F_{00}} \bigg) \alpha_n + 
\bigg(\frac{W}{2H}+\frac{WK}{F_{00}}\bigg)
\big(\beta_n + 2\gamma_n\big) \bigg] ,
\end{equation} 
\begin{equation}\label{16b}
\delta_{n+1} =
\frac{a^2}{2(n+3)} \bigg[ 
\dot\alpha_n+\dot\beta_n+2\dot\gamma_n +  
\bigg(6H+\frac{\dot F_{00}}{F_{00}}\bigg)\alpha_n +
\frac{WK}{F_{00}}
\big(\beta_n+2\gamma_n\big) \bigg] ,
\end{equation} 
\begin{equation}\label{16c}
\alpha_{n+2} = 
\frac{1}{n+2}\,\frac{1}{H} \left\{ \Big[\beta_{n+2} - 
\big(n+3\big)\gamma_{n+2}\Big]_{\!,0} - 
W\delta_{n+1} \right\} ,
\end{equation} 
\begin{equation}\label{16d}
\Big(\frac{n}{2}+1\Big)\alpha_{n+2} + 
\Big(\frac{n}{2}+3\Big)\beta_{n+2} -
\big(n+4\big)\gamma_{n+2}  =
\dot\delta_{n+1} + \big(H + K\big)
\delta_{n+1} \,.                  
\end{equation} 
\end{subequations}
\end{widetext}
The full equivalence with the initial set of field equations is
obtained when the recurrent relations (\ref{16}) are supplemented with
\begin{equation}\label{17}
C_{-3} = 0 \,.  
\end{equation} 
Now, Eqs.\ (\ref{16}) and (\ref{17}) represent the full set of
differential equations that govern the dynamics of small, spherically
symmetric perturbations of the metric. It should be emphasized that
the described procedure is applicable to any cosmological background.
Indeed, the scale factor $a$, and the potential $W$ have not been
specified so far.

\section{Solution}\label{Sec4}

The Eqs.\ (\ref{16}) and (\ref{17}) of the preceding section hold true
for any choice of the scale factor $a$, and the potential $W$.
Unfortunately, the interesting choices, such as inflationary or
bouncing cosmologies, turn out to be quite involved. For this reason,
I shall turn to the commonly accepted concept that, whatever type of
cosmology is considered, its late time behavior should be that of the
$\Lambda$CDM model. In what follows, I shall be interested in the
vicinity of the present epoch. This leads me to make a simple choice
\begin{equation}\label{18}
a = e^{\omega t} \,, \quad 
W = - \frac{\omega^2}{a^2} \,,
\end{equation} 
where $\omega$ is a constant with the dimension of mass. What one
should have in mind is that the above exponential law is just the late
time behavior of a more general $a(t)$. In particular, $a(t)$ could
have an inflationary period like in
$$
a = \frac{e^{\omega t}}{1+e^{-67\omega t}} \,,
$$
or a bounce as in
$$
a = \sqrt[3]{\ds e^{\omega t}-\omega t} \,.
$$
In both these examples, the late time behavior of $a(t)$ is
exponential, as expected from the cosmology of the present epoch. The
definition Eq.\ (\ref{18}) is straightforwardly checked to satisfy the
regularity and stability conditions (\ref{6.2}) and (\ref{6.4}).

The simple geometric sigma model defined by Eq.\ (\ref{18})
considerably simplifies the recurrent relations of the preceding
section. For one thing, the Hubble parameter becomes a constant, as
the exponential law of the scale factor implies $H=\omega$. In
accordance with the measured value of the Hubble parameter, the
numerical value of this constant must be
\begin{equation}\label{18.5}
\omega = 0.75 \times 10^{-10} \ {\rm yr}^{-1} \equiv
1.6 \times 10^{-33} \ {\rm eV} \,. 
\end{equation} 
The choice of the potential $W = - \omega^2/a^2$ , on the other hand,
implies $K=0$. With these simplifications, Eqs.\ (\ref{16}) and
(\ref{17}) take the form
\begin{subequations}\label{19}
\begin{equation}\label{19a}
\begin{array}{rl}
&\ds \dot\alpha_n + \omega\alpha_n -  
\frac{\omega}{2a^2} \big( \alpha_n + 
\beta_n + 2\gamma_n \big) =                                        \\
&\ds \frac{n+3}{\omega a^2}\big[\beta_{n+2} - 
(n+3)\gamma_{n+2}\big] ,                                            \\[0.4ex]
\end{array}
\end{equation} 
\begin{equation}\label{19b}
\begin{array}{rl}
&\ds \dot\alpha_n+\dot\beta_n+2\dot\gamma_n +  
4\omega\alpha_n = 2\frac{n+3}{a^2}\delta_{n+1} \,,   \\[0.4ex]
\end{array}
\end{equation} 
\begin{equation}\label{19c}
\begin{array}{rl}
&\ds \dot\beta_{n+2} - (n+3)\dot\gamma_{n+2} + 
\frac{\omega^2}{a^2}\delta_{n+1} =
(n+2)\omega\alpha_{n+2} \,,                                         \\[0.4ex]  
\end{array}
\end{equation} 
\begin{equation}\label{19d}
\begin{array}{rl}
&\ds \dot\delta_{n+1} + \omega \delta_{n+1} -
2\big[\beta_{n+2}-(n+3)\gamma_{n+2}\big] =               \\
&\ds \frac{n+2}{2}\big(\alpha_{n+2}+\beta_{n+2} +
2\gamma_{n+2}\big)  \,,                                                 \\[0.4ex]  
\end{array}
\end{equation} 
\begin{equation}\label{19e}
\begin{array}{rl}
&\ds a^2\left[\big(\beta_{-3}-\gamma_{-3}\big)_{\!,00} +
3\omega\big(\beta_{-3}-\gamma_{-3}\big)_{\!,0}\right] +    \\
&\ds 2\omega^2\big(\beta_{-3}-\gamma_{-3}\big) +   
6\left(\dot\delta_{-2} + \omega \delta_{-2}\right) =              \\
&\ds 3\big(\beta_{-1} - \alpha_{-1}\big) \,.  
\end{array}
\end{equation} 
\end{subequations}
In the next section, I shall demonstrate that geodesic equation in the
nonrelativistic approximation does not depend on $\lambda_1$ and
$\rho_1$. Therefore, the only coefficients needed for the evaluation
of the gravitational acceleration are $\alpha_n$ and $\delta_n$. The
respective recurrent relations are obtained as appropriate linear
combinations of Eqs.\ (\ref{19}). One finds
\begin{equation}\label{20}
\ddot\delta_{n+1} + \omega\dot\delta_{n+1} +
2\frac{\omega^2}{a^2}\delta_{n+1} =
\frac{(n+2)(n+5)}{a^2}\delta_{n+3} \,,
\end{equation} 
\begin{equation}\label{21}
\ddot\alpha_n + 3\omega\dot\alpha_n +
2\omega^2\Big(1+\frac{1}{a^2}\Big)\alpha_n =
\frac{(n+2)(n+3)}{a^2}\alpha_{n+2} \,.
\end{equation}
The first equation is obtained by differentiating Eq.\ (\ref{19d}),
and by subsequent replacement of $\dot\beta_{n+2} -
(n+3)\dot\gamma_{n+2}$ and $\dot\alpha_{n+2}+\dot\beta_{n+2}+
2\dot\gamma_{n+2}$ from Eqs.\ (\ref{19c}) and (\ref{19d}),
respectively. The second equation is obtained by the substitution of
$\alpha_n + \beta_n + 2\gamma_n$ from Eq.\ (\ref{19a}) into Eq.\
(\ref{19b}), and subsequent elimination of $\dot\beta_{n+2} -
(n+3)\dot\gamma_{n+2}$ with the help of Eq.\ (\ref{19c}). In what
follows, every solution of Eqs.\ (\ref{20}) and (\ref{21}) will be
checked for its consistency with the complete set of equations
(\ref{19}).

Let me now solve the above recurrent relations. In the first step, the
full set of equations is divided into two mutually independent groups.
The first group consists of all the equations whose index $n$ is odd.
The second group is characterized by even $n$. The solutions of these
two groups do not mix with each other.

\subsection{Odd values of {\boldmath$n$}}\label{Sec4a}

Eqs.\ (\ref{19}), (\ref{20}) and (\ref{21}) are most easily solved if the
infinite series Eq.\ (\ref{11}) is truncated at some negative value of
the index $n$. In this subsection, I shall use the ansatz
\begin{equation}\label{22}
\alpha_{2k-1} = \beta_{2k-1} = 
\gamma_{2k-1} = \delta_{2k} = 0 \quad \forall \ k\leq -1 \,.
\end{equation}
This ansatz identically satisfies Eqs.\ (\ref{19}), (\ref{20}) and
(\ref{21}) for all odd $n\leq -5$. Let us see what happens when $n\geq
-3$. If we start with $n=-3$, the following solution is found. From
Eq.\ (\ref{19c}) if follows $\dot\beta_{-1} + \omega\alpha_{-1} = 0$,
whereas Eq.\ (\ref{19e}) yields $\alpha_{-1}=\beta_{-1}$. These two
equations lead to
\begin{equation}\label{23}
\alpha_{-1}=\beta_{-1}=\frac{\ell}{a} \,,
\end{equation}
where $\ell$ is a free constant with the dimension of length. Finally,
Eq.\ (\ref{19d}) tells us that $\gamma_{-1} = \beta_{-1} =
\alpha_{-1}$, while Eq.\ (\ref{20}) gives
\begin{equation}\label{24}
\delta_{2k} = 0 \quad \forall \ k \,.
\end{equation}
The same procedure is readily applied to higher values of $n$. The
final result is
\begin{equation}\label{25}
\gamma_{2k-1} = \beta_{2k-1} = \alpha_{2k-1} =
\frac{\big(\sqrt{2}\omega\big)^{2k}}{(2k)!} \frac{\ell}{a}
\quad \forall \ k\geq 0 \,.
\end{equation}
The coefficients not included in Eqs.\ (\ref{24}) and (\ref{25}) are
determined from the ansatz Eq.\ (\ref{22}). The needed metric
components are obtained straightforwardly. Specifically,
\begin{equation}\label{26}
\left(\mu_1\right)_{\rm odd} = 
\frac{\ell}{ar} \sum_{n=0}^{\infty}
\frac{\big(\sqrt{2}\omega r\big)^{2n}}{(2n)!} \,,\quad
\left(\nu_1\right)_{\rm even} =0 \,,
\end{equation}
where $(\mu_1)_{\rm odd}$ stands for the odd part of $\mu_1$, and
$(\nu_1)_{\rm even}$ denotes even part of $\nu_1$. The variables
$\lambda_1$ and $\rho_1$ are obtained straightforwardly, but I choose
not to display them here. This is because the evaluation of the
nonrelativistic gravitational acceleration, which is the main
objective of this paper, turns out not to depend on these two
variables.

\subsection{Even values of {\boldmath$n$}}\label{Sec4b}

Let me truncate the series Eq.\ (\ref{11}) by applying the ansatz
\begin{equation}\label{27}
\alpha_{2k} = \beta_{2k} = 
\gamma_{2k} = \delta_{2k-1} = 0 \quad \forall \ k\geq 2 \,.
\end{equation}
It is immediately seen that Eqs.\ (\ref{19}), (\ref{20}) and
(\ref{21}) are identically satisfied for all even $n\geq 4$. For other
even values of $n$, the following holds true. All $\beta_n$ and
$\gamma_n$ are uniquely determined in terms of $\alpha_n$ and
$\delta_{n-1}$, provided the latter are solutions of Eqs.\ (\ref{20})
and (\ref{21}). Moreover, this holds true for every such pair of
solutions to Eqs.\ (\ref{20}) and (\ref{21}). The complete set of
equations (\ref{19}) does not bring any further restrictions.

With these preliminaries, the needed metric components become
\begin{equation}\label{28}
\begin{array}{rl}
&\ds \big(\mu_1\big)_{\rm even} =
\alpha_2 r^2 + \alpha_0 + 
{\cal O}\Big(\frac{1}{r^2}\Big) ,                       \\ [1ex] 
&\ds \big(\nu_1\big)_{\rm odd} =
\delta_1 r + \frac{\delta_{-1}}{r} + 
{\cal O}\Big(\frac{1}{r^3}\Big) , 
\end{array}
\end{equation}
where the coefficients $\alpha_2$, $\alpha_0$, $\delta_1$,
$\delta_{-1}$ are solutions of the corresponding Eqs.\ (\ref{20}) and
(\ref{21}). With the help of the ansatz Eq.\ (\ref{27}), the general
solution is found to have the form
\begin{equation}\label{29}
\begin{array}{rl}
&\ds \alpha_2 = 
\frac{1}{a}\Big[ c'_2\cos\Big(\frac{\sqrt{2}}{a}\Big) +
c''_2\sin\Big(\frac{\sqrt{2}}{a}\Big)\Big] ,                        \\ [1ex]  
&\ds \delta_1 = 
c'_1\cos\Big(\frac{\sqrt{2}}{a}\Big) +
c''_1\sin\Big(\frac{\sqrt{2}}{a}\Big) ,                               \\ [1ex] 
&\ds \delta_{-1} = 
c'_{-1}\cos\Big(\frac{\sqrt{2}}{a}\Big) +
c''_{-1}\sin\Big(\frac{\sqrt{2}}{a}\Big) , 
\end{array}
\end{equation}
where $c'_n$ and $c''_n$ are free integration constants. The
coefficient $\alpha_0$ has deliberately been omitted because it does
not appear in the expression for the gravitational acceleration. This
will become clear in the next section, where the formula for the
gravitational acceleration will be derived from the nonrelativistic
geodesic equation.

\section{Gravitational acceleration}\label{Sec5}

The formula for gravitational acceleration is derived from the geodesic equation
$$
\frac{du^{\mu}}{ds} + \Gamma^{\mu}{}_{\nu\rho}
u^{\nu}u^{\rho} = 0 \,,
$$
where $u^{\mu} \equiv dx^{\mu}/ds$. As typical astronomical velocities
are much smaller than the speed of light, we shall work in the
nonrelativistic approximation. Then, the geodesic equation for the
metric Eq.\ (\ref{7}) is brought to the form
\begin{equation}\label{30}
\frac{dv^{\alpha}}{dt} + 
\Gamma^{\alpha}{}_{00} = {\cal O}_2 \,, 
\end{equation} 
where $v^{\alpha} \equiv dx^{\alpha}/dt$, and ${\cal O}_2$ denotes
terms of second order in velocities and metric perturbations.
(Precisely, $\mu_1$, $\nu_1$, $\lambda_1$, $\rho_1$ and $v^{\alpha}$
are all considered ${\cal O}_1$ terms.) The component
$\Gamma^{\alpha}{}_{00}$ is straightforwardly found from
Eqs.\ (\ref{7}) and (\ref{9}). Then, Eq.\ (\ref{30}) takes the form
\begin{equation}\label{31}
\frac{d\vec v}{dt} =
\frac{1}{2a^2}\big(\mu'_1 - 2\dot\nu_1\big)
\frac{\vec r}{r} + {\cal O}_2 \,.
\end{equation} 
Physical acceleration is obtained by using the physical distance
$$
dr_{\rm phys} = a(t) dr \,,
$$
which should be integrated out to give the global physical distance
$r_{\rm phys}$. As meaningful notion of global distance is
known to require static geometry, we shall restrict to small time
intervals in which $a(t)$ remains practically unchanged. Then, one
finds
$$
r_{\rm phys} \approx a(t_*)\,r \,,\quad
\vec v_{\rm phys}\approx  a(t_*)\,\vec v \,,
$$
for all $t$ in the vicinity of $t_*$. The time $t_*$ in the above
formulas is a fixed time, which can be thought of as the cosmic time
the observed astronomical object lives in. In all final expressions, I
shall replace $t_*$ with more common $t$. Then, the magnitude of the
physical acceleration $\vec g \equiv d \vec v_{\rm phys}/dt$ takes the
form
\begin{equation}\label{32}
g = \frac{1}{2a}\big(\mu'_1 - 2\dot\nu_1\big) \,.
\end{equation} 
The direction of $\vec g$ coincides with that of $\vec r$. This means
that negative $g$ stands for an attractive force, whereas positive $g$
is repulsive. The final form of the gravitational acceleration is
obtained when $\mu_1 \equiv (\mu_1)_{\rm odd} + (\mu_1)_{\rm even}$,
and $\nu_1 \equiv (\nu_1)_{\rm odd} + (\nu_1)_{\rm even}$ are derived
from Eqs.\ (\ref{26}), (\ref{28}) and (\ref{29}), and then substituted
into Eq.\ (\ref{32}). This way, one finds
\begin{equation}\label{33}
g = \frac{\omega^2\ell}{2} \left[ -\frac{1}{x^2} -
\frac{Q(t)}{x} + P(t)x + J(t,x) \right] + 
{\cal O}\Big(\frac{1}{x^3}\Big) ,
\end{equation} 
where
\begin{equation}\label{34}
x \equiv \omega r a
\end{equation} 
is the physical distance in units of the Hubble length. The time
dependent coefficients $Q(t)$ and $P(t)$ are derived from Eqs.\
(\ref{29}). By an appropriate redefinition of the free integration
constants $c'_n$ and $c''_n$, they are brought to the form
\begin{equation}\label{35}
Q = Q_0\,\frac{\sin\left(\frac{\sqrt{2}}{a}+\theta\right)}
{\sin\left(\frac{\sqrt{2}}{a_0}+\theta\right)}\,\frac{a_0}{a} 
\end{equation} 
and
\begin{equation}\label{36}
P = P_0\, \frac{\sin\left(\frac{\sqrt{2}}{a}+\phi\right)}
{\sin\left(\frac{\sqrt{2}}{a_0}+\phi\right)}
\Big(\frac{a_0}{a}\Big)^{\!3} ,
\end{equation} 
where $Q_0$, $P_0$, $\theta$ and $\phi$ are the redefined integration
constants, and $a_0 \equiv a(t_0)$ is the value of the scale factor at
the present epoch $t=t_0$. The coefficient $J(t,x)$ has the form
\begin{equation}\label{37}
J = \frac{1}{x^2} \sum_{n=1}^{\infty}
\frac{2n-1}{(2n)!}\Big(\frac{\sqrt{2}x}{a}\Big)^{\!2n} ,
\end{equation} 
which is the power expansion of the function
\begin{equation}\label{38}
J = \frac{1}{x^2}\Big[1-\cosh\Big(\frac{\sqrt{2}x}{a}\Big) +
\frac{\sqrt{2}x}{a}\sinh\Big(\frac{\sqrt{2}x}{a}\Big)\Big] . 
\end{equation} 
It is seen that $J(t,x)$ is always positive, exponentially increasing
function of $x$. Exponential corrections to the Newtonian force are
not new in scientific literature. For example, in Ref.\ \cite{18b},
such corrections are obtained from the effective quantum gravity
theory. When compared to the present result, an important difference
is noted. While modified gravitational force of Ref.\ \cite{18b} is
everywhere decreasing function of distance, the term Eq.\ (\ref{38})
gives an exponentially increasing contribution to the gravitational
acceleration Eq.\ (\ref{33}).

The analysis of Eq.\ (\ref{33}) shows that the gravitational
acceleration is a sum of four major contributions, which one by one,
become dominant as the distance from the gravitational source grows.
In what follows, this fact will be used for making a comparison with
the known observational data. To this end, I shall make use of two
well established theories that have already been verified to correctly
interpret astronomical measurements. These are the phenomenological
MOND theory that gives a satisfactory description of galaxies, and
$\Lambda$CDM model that correctly explains late time cosmology. No
direct comparison with raw astronomical measurements will be done.
Nevertheless, the comparison with MOND and $\Lambda$CDM will help us
determine some of the remaining free integration constants.

\subsection{Comparison with MOND}\label{Sec5a}

Let me start with with the analysis at short distances. In this case,
the gravitational acceleration Eq.\ (\ref{33}) is dominated by its
first term, which reduces to the Newtonian term
\begin{equation}\label{39}
g_N = -\frac{GM}{r^2_{\rm phys}} 
\end{equation} 
if the integration constant $\ell$ is chosen in the form 
\begin{equation}\label{40}
\ell = 2GM \,.
\end{equation} 
With $M$ representing the source mass, and $G$ the gravitational
constant, the constant $\ell$ becomes the Schwarzschild radius of the
pointlike source.

At slightly larger distances, the second term in Eq.\ (\ref{33}) comes
into play. This kind of term has already been suggested in literature
in connection with the problem of flat galactic curves. One of the
most cited phenomenological models is Milgrom's modified Newtonian
dynamics, commonly referred to as MOND \cite{9,10,10a,11,12,13,14,15}.
It succeeded in explaining flat galactic curves by employing a
modification analogous to the second term of Eq.\ (\ref{33}). In the
present cosmic epoch, the MOND gravitational acceleration reads
\begin{equation}\label{41}
\left(g\right)_{MOND} = -\frac{GM}{r^2_{\rm phys}} -
\frac{\sqrt{GM g_0}}{r_{\rm phys}} \,,
\end{equation} 
where
$$
g_0 = 1.26 \times 10^{-11}\ {\rm yr}^{-1}
$$
is the MOND universal acceleration constant. The comparison with the
present time form of Eq.\ (\ref{33}) then gives us the value of the
integration constant $Q_0$. Precisely,
\begin{equation}\label{42}
Q_0 = \sqrt{\frac{2 g_0}{\omega^2\ell}} \,.
\end{equation} 
The integration constants $\ell$ and $Q_0$, as given by Eqs.\
(\ref{40}) and (\ref{42}), enable the present time value of Eq.\
(\ref{33}) to have the exact MOND behavior for a wide range of
distances. (At very large distances, the gravitational acceleration
leaves the MOND regime in favor of the repulsive force that governs
the accelerated expansion of the Universe.) As for the time dependence
of the gravitational acceleration, it is seen from Eq.\ (\ref{35})
that it has oscillatory character. The period of these oscillations is
given by the formula
$$
\Delta \left(\omega t\right) = \ln \left(1+\pi\sqrt{2}\,a\right) ,
$$
which is directly read from $Q(t)$. Obviously, the oscillations become
more rapid as we go to the past. In the vicinity of the present epoch,
on the other hand, we have
$$
\Delta\left(\omega t\right) = 
\ln\left(1+\pi\sqrt{2}\,a_0\right) > 1.7
$$
whenever $t_0>0$. This time interval lies far beyond the observable
Universe, so that the oscillatory nature of the gravitational force is
practically undetectable. This is a consequence of the condition
$$
t_0 > 0 \,,
$$
which is easily justified using the experimental bound on the graviton
mass. Indeed, the condition $t_0>0$ implies that the graviton mass, as
defined by Eq.\ (\ref{6.6}), obeys the inequality
$$
m_g < 2.3 \times 10^{-33}\ {\rm eV} \,,
$$
which is more than ten orders of magnitude smaller than the
experimental bound reported by the LIGO experiment \cite{27}. The
condition $t_0>0$ is also supported by other estimates of the graviton
mass that can be found in literature \cite{28}.

\begin{figure}[htb]
\begin{center}
\includegraphics[height=4cm]{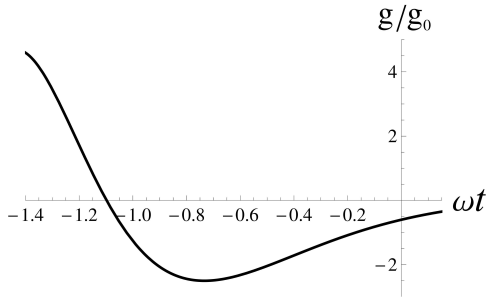}
\end{center}
\vspace*{-.5cm}
\caption{Time dependence of gravitational acceleration.\label{f1} }
\end{figure}

Finally, let me say something about $\theta$ dependence of the
gravitational acceleration in MOND regime. First, it is seen that at
$t=t_0$, the gravitational acceleration does not depend on $\theta$,
at all. At $t<t_0$, however, the $\theta$ dependence becomes quite
significant. In particular, the value of $\theta$ determines if the
gravitational force in the vicinity of $t=t_0$ increases or decreases
with time. To illustrate the form of time dependence that $g(t,x)$ can
have, let me consider the simple example
$$
t_0 = 0 \,, \quad \theta = -\frac{\pi}{3} \,, \quad
\phi = 0 \,,
$$
and apply it to the gravitational source 
$$
\ell = 0.08\ \,{\rm ly} \,, \quad x = 10^{-5} \,. 
$$
(These values of $\ell$ and $x$ correspond to the mass and radius of
Milky Way.) The graph of the function $g(t,x)$ is depicted in Fig.\ \ref{f1}. 
As one can see, the gravitational acceleration is weaker now than it
used to be in the recent past. In the distant past, on the other hand,
the gravitational force begins to oscillate. One should have in mind,
however, that this picture drastically changes if the present time
$t_0$ takes much larger values. Then, the period of oscillations
grows beyond physical detection. In particular, the limit $t_0 \to
\infty$ turns the oscillatory behavior into a simple exponential law.
Precisely, as one goes to the past, the attractive gravitational
acceleration experiences exponentially fast growth.

\subsection{Comparison with {\boldmath$\Lambda$}CDM}\label{Sec5b}

The gravitational acceleration of a pointlike source in the $\Lambda$CDM 
background has the form
\begin{equation}\label{43}
(g)_{\Lambda CDM} = -\frac{GM}{r^2_{\rm phys}} + 
\omega^2 r_{\rm phys} \,.
\end{equation} 
Unfortunately, our expression Eq.\ (\ref{33}) can never fully reduce to
this form. What one can do is to make use of the fact that there is a
range of distances for which the third term of Eq.\ (\ref{33}), and
the second term of Eq.\ (\ref{43}) become dominant terms of their
respective expressions. The integration constant $P_0$ is then
determined from the requirement that these two dominant terms coincide
at the present epoch. This leads to
\begin{equation}\label{44}
P_0 = \frac{2}{\omega\ell} \,.
\end{equation} 
The constant $P_0$, as defined by Eq.\ (\ref{44}), ensures that the
accelerated expansion predicted in this paper coincides with that of
$\Lambda$CDM model.

Let me now calculate the turnaround radius of an arbitrarily chosen
gravitational source, and compare it with the corresponding
$\Lambda$CDM expression. The turnaround radius is defined as the
distance from the pointlike source at which gravitational force drops
to zero. In the type of theories we consider, the notion of turnaround
radius is always well defined. Indeed, the gravitational force is
attractive at small distances, whereas at large distances it becomes
repulsive. Thus, there must exist the distance at which the
gravitational acceleration takes zero value. It is calculated from the
equation $g=0$, which reduces to
$$
-\frac{1}{x^2} - \frac{Q}{x} + Px + J = 0 \,,
$$
as seen from Eq.\ (\ref{33}). I have already mentioned earlier that
the last term in the above expression becomes important only at
extremely large distances. Such large radii, however, are not met in
the contemporary astronomical measurements. As a consequence, the term
$J$ is shown to have a negligible influence on the value of $g$. The
present epoch turnaround radius is then found by solving the equation
\begin{equation}\label{45}
P_0\, x^3 - Q_0\, x - 1 = 0 \,.
\end{equation} 
The general solution of cubic algebraic equations is well known, so
that one straightforwardly obtains
\begin{equation}\label{46}
x_t = \left(2P_0\right)^{-\frac 13} \left[
\left(1+\sqrt{1-\sigma}\right)^{\frac 13} +
\left(1-\sqrt{1-\sigma}\right)^{\frac 13} \right] ,
\end{equation} 
where $x_t$ denotes turnaround radius, and
\begin{equation}\label{47}
\sigma \equiv \frac{4}{P_0}\!\left(\frac{Q_0}{3}\right)^{\!3} .
\end{equation} 
To estimate the range of values of $\sigma$, let me make use of the
fact that no astronomical object in the observable Universe has mass
larger than $\ell = 10^5$ ly. This leads to $\sigma \gtrsim 5.25$, so
that both square roots in Eq.\ (\ref{46}) are imaginary. With this,
the turnaround radius $x_t$ takes the form
\begin{equation}\label{48}
x_t = \left(2P_0\right)^{-\frac 13} \left[
\left(1+i\sqrt{\sigma-1}\right)^{\frac 13}+{\rm c.c.}\,\right] .
\end{equation} 
The complicated expression Eq.\ (\ref{48}) is simplified as follows.
One starts with
$$
1+i\sqrt{\sigma-1} \equiv \sqrt{\sigma} e^{i\varphi} ,
$$
where $\varphi \equiv \arctan\sqrt{\sigma-1}$. Then, the turnaround
radius $x_t$ is straightforwardly brought to the form
$$
x_t \approx 0.62 \cdot \cos\frac{\varphi}{3} \cdot 
\left(\omega\ell\right)^{\frac 14} . 
$$
This expression is further simplified by noticing that
$\cos\frac{\varphi}{3}$ remains practically unchanged in the interval
$\ell\in (0, 10^5)$ ly. Indeed, the constraint $\ell<10^5$ ly implies
$\sigma \gtrsim 5.25$, which yields $1.12 \lesssim \varphi < \pi/2$.
As a consequence, $0.87 \lesssim \cos\frac{\varphi}{3} \lesssim 0.93$,
so that $\cos\frac{\varphi}{3}$ can approximately be considered a
constant. Specifically, $\cos\frac{\varphi}{3} \approx 0.90 \pm 0.03$
for all $\ell<10^5$ ly. The turnaround radius is then rewritten as
\begin{equation}\label{49}
x_t \approx \left( 0.56 \pm 0.02 \right) \cdot 
\left( \omega\ell \right)^{\frac 14} .
\end{equation} 
This formula holds true for all galaxy clusters and superclusters in the observable Universe. The corresponding $\Lambda$CDM expression is
obtained by solving the equation $(g)_{\Lambda{\rm CDM}} = 0$. It
results in
\begin{equation}\label{50}
(x_t)_{\Lambda{\rm CDM}} \approx 0.79 \cdot 
\left( \omega\ell \right)^{\frac 13}. 
\end{equation} 
The comparison of Eq.\ (\ref{49}) with Eq.\ (\ref{50}) tells us that
our turnaround radius does not agree with that of $\Lambda$CDM model.
This is a consequence of the presence of MOND term in our expression
for gravitational acceleration. One should have in mind though that
there is still a possibility to change the form of the turnaround
radius by making a different choice of the integration constant $P_0$.
While this can ensure that the two turnaround radii become compatible,
the accelerated expansion of the Universe will inevitably loose its
$\Lambda$CDM form.

\section{Concluding remarks}\label{Sec6}

I have considered in this paper a class of cosmological models based
on geometric sigma models with four scalar fields. These models have
already been examined in Ref.\ \cite{18}, where their regularity and
stability have been proven. In this work, I search for spherically
symmetric solutions, with the idea to check how additional four
degrees of freedom modify Newtonian gravitational law.

The main result of my calculations is given by Eq.\ (\ref{33}), which
represents gravitational acceleration of a pointlike source in de
Sitter background. In fact, the obtained result refers to the late
time behavior of any cosmology with $\Lambda$CDM limit. There are four
major terms in Eq.\ (\ref{33}), which one by one, become dominant as
the distance from the gravitational source grows. At short distances,
the dominant contribution comes from the Newtonian term. As the
distance grows, the dominant role is taken by the familiar MOND term.
At even larger distances, the leading role is carried by the
$\Lambda$CDM term which is responsible for the accelerated expansion
of the Universe. Finally, the fourth term provides an extra repulsive
acceleration that grows exponentially fast with distance. This term
becomes significant only at extremely large distances that go beyond
the observable Universe. As such, it is effectively neglected. The
analysis has been done in a weak field approximation, with the help of
an additional assumption that restrains the overall generality.
Precisely, the field equations are solved with the help of the ansatz
Eqs.\ (\ref{22}) and (\ref{27}) that basically truncated the infinite
power series Eq.\ (\ref{11}). Owing to this, the linearized field
equations have been successfully solved. One should keep in mind,
however, that the obtained solution is not as general as one would
ideally like to have.

Another simplification used in this paper is the abandonment of the
fourth term in Eq.\ (\ref{33}). It has been explained that, in all
observationally interesting situations, this term is too small. Let me
clarify this statement. The term $J$ is positive, exponentially
increasing function of $x$, which obviously becomes dominant for large
enough $x$. In practice, however, the value of $x$ is bounded by the
fact that the largest observed astronomical object has diameter of the
order of $10^8$ ly. As a consequence, the distance $x$ is constrained
by the inequality $x \lesssim 10^{-2}$. The cosmic time is also
constrained. Indeed, the observable history of the Universe is defined
by the finite interval $-1\lesssim \omega (t-t_0) \leq 0$. With these
restrictions, the argument of the function $J$ is found to satisfy
$\sqrt{2}x/a \lesssim 0.04/a_0$, which straightforwardly leads to
$$
J \lesssim 7  
$$
for all $t_0>0$. For higher values of the present time $t_0$, the term
$J$ is constrained even more. Let me now compare $J$ with other terms
in Eq.\ (\ref{33}). The term $1/x^2$ is estimated with the help of the
restriction $x \lesssim 10^{-2}$. It immediately gives
$$
\frac{1}{x^2} \gtrsim 10^4 \,,
$$
which tells us that $J \ll 1/x^2$. The term $Q/x$ is estimated with
the help of three observational restrictions. The first two are $x
\lesssim 10^{-2}$ and $-1\lesssim \omega (t-t_0) \leq 0$, while the
third comes from the observation that no astronomical object in the
observed Universe has mass larger than $\ell\sim 10^5$ ly. These three
restrictions yield
$$
\frac{Q}{x} \gtrsim 10^5 \,,
$$
and consequently, $J \ll Q/x$. Finally, let me estimate the term $Px$.
It is immediately seen that $Px$ can be arbitrarily small if we
restrict to small distances from the gravitational source. Notice,
however, that the observed galaxies, galaxy clusters and superclusters
are not pointlike objects. Instead, they have nonzero radii, which are
related to their masses. The needed mass is the one which is
distributed below the chosen distance $x$. A rough estimation of how
$\ell$ is related to $x$ can be obtained in the spherically symmetric
approximation in which matter density is considered constant. This
assumption immediately leads to $\ell \propto x^3$, and consequently,
$P_0\,x \propto 1/x^2$. Thus, the term $P_0\,x$ is bounded from below
by the fact that $x \lesssim 10^{-2}$. The estimation of $P x$ is
obtained when the restrictions $-1\lesssim \omega (t-t_0) \leq 0$ and
$\ell \lesssim 10^5$ ly are taken into account. One straightforwardly
finds
$$
P x \gtrsim 10^3 \,,
$$
so that $J \ll P x$. As we can see, $J$ is indeed negligible when
compared to the other three terms in Eq.\ (\ref{33}). The obtained
estimation holds true when the present time $t_0$ obeys the inequality
$t_0>0$. It applies to all the astronomical objects in the observable
Universe.

To summarize, I have shown in this paper that observationally
justified modifications of Newtonian gravity do not have to be imposed
by hand. Instead, they can be found in already existing cosmological
models. Specifically, I have examined a class of geometric sigma
models whose late time behavior reduces to that of $\Lambda$CDM model.
As it turns out, each of these models accommodates a spherically
symmetric solution with the required MOND and $\Lambda$CDM
modifications. Irrespective of this success, the present work is far
from being complete. What remains to be done is to find the physical
interpretation of the remaining physical degrees of freedom. This
task, however, lies far beyond the scope of this paper.

\begin{acknowledgments}
This work is supported by the Serbian Ministry of Education, Science
and Technological Development, under Contract No. $171031$.
\end{acknowledgments}


\begin{thebibliography}{99}

\bibitem{1}
J. C. Kapteyn, Astrophys. J. \textbf{55}, 302 (1922).

\bibitem{2}
J. H. Oort, Bulletin of the Astronomical Institutes of the Netherlands
\textbf{6}, 249 (1932).

\bibitem{3}
F. Zwicky, Helvetica Physica Acta \textbf{6}, 110 (1933).

\bibitem{4}
F. Zwicky, Astrophys. J. \textbf{86}, 217 (1937).

\bibitem{5}
H.W. Babcock, Lick Observatory Bulletins \textbf{19}, 41 (1939).

\bibitem{6}
V. C. Rubin and W. K. Ford Jr., Astrophys. J. \textbf{159}, 379 (1970).

\bibitem{7}
V. C. Rubin, N. Thonnard and W. K. Ford Jr., Astrophys. J. 
\textbf{238}, 471 (1980).

\bibitem{8}
S. Capozziello and M. De Laurentis, Physics Reports \textbf{509}, 167, (2011).

\bibitem{16}
M. Vasilic, Class. Quant. Grav. \textbf{15}, 29 (1998).

\bibitem{18}
M. Vasilic, Phys. Rev. D \textbf{95}, 123506 (2017).

\bibitem{18x}
V. Borka Jovanovic, S. Capozziello, P. Jovanovic and D. Borka,
Phys. Dark Univ. \textbf{14}, 73 (2016).

\bibitem{18y}
S. Capozziello, P. Jovanovic, V. Borka Jovanovic and D. Borka,
JCAP \textbf{1706}, 044 (2017). 

\bibitem{18a}
M. Vasilic, Class. Quantum Grav. \textbf{35}, 175015 (2018).

\bibitem{18b}
Xavier Calmet, Salvatore Capozziello and Daniel Pryer,
Eur. Phys. J. C \textbf{77}, 589 (2017).

\bibitem{9}
M. Milgrom, Astrophys. J. \textbf{270}, 365 (1983).

\bibitem{10}
M. Milgrom, Astrophys. J. \textbf{270}, 371 (1983).

\bibitem{10a}
B. Famaey and S. McGaugh, Living Rev. Relativity \textbf{15}, 10 (2012). 

\bibitem{11}
R. H. Sanders and S. S. McGaugh, Ann. Rev. Astron. Astrophys. 
\textbf{40}, 263 (2002).

\bibitem{12}
J.D. Bekenstein, Contemp. Phys. \textbf{47}, 387 (2006).

\bibitem{13}
C. Skordis, Class. Quant. Grav. \textbf{26}, 143001 (2009).

\bibitem{14}
S. S. McGaugh and W. J. G. De Blok, Astrophys. J. \textbf{499}, 66 (1998).

\bibitem{15}
S. S. McGaugh, Phys. Rev. Lett. \textbf{106}, 121303 (2011).

\bibitem{27}
B. P. Abbott et al. [LIGO Scientific Collaboration, Virgo Collaboration],
Phys. Rev. Lett. \textbf{116}, 061102 (2016);
\textbf{116}, 221101 (2016); \textbf{116}, 241103 (2016);
Phys. Rev. X \textbf{6}, 041015 (2016).

\bibitem{28}
C. de Rham, J. T. Deskins, A. J. Tolley and S. Y. Zhou,
Rev. Mod. Phys. \textbf{89}, 025004 (2017).




\end{thebibliography}
\end{document}